\documentclass[manuscript]{acmart}

\AtBeginDocument{%
  }

\acmConference[xxx '25]{}
\newcommand{\yun}[1]{\out{{\small\textcolor{teal}{\bf [*** Yun: #1]}}}}

\newcommand{\si}[1]{\out{{\small\textcolor{red}{\bf [*** Si: #1]}}}}

\usepackage{graphicx}
\usepackage{multirow}
\usepackage{makecell}
\begin{document}

\title[Motion-Driven Design]{Redesigning Educational Videos for Deaf and Hard-of-Hearing Learners}



\author{Si Chen}
\email{sic3@illiois.edu}
\orcid{0000-0002-0640-6883}
\affiliation{%
  \institution{School of Information Sciences, University of Illinois Urbana-Champaign}
  \city{Champaign}
  \state{Illinois}
  \country{USA}
  \postcode{61802}
}

\author{Haocong Cheng}
\email{haocong2@illinois.edu}
\affiliation{
  \institution{School of Information Sciences, University of Illinois Urbana-Champaign}
  \city{Champaign}
  \state{Illinois}
  \country{USA}
}

\author{Suzy Su}
\email{xiaoyus4@illinois.edu}
\affiliation{
  \institution{School of Information Sciences, University of Illinois Urbana-Champaign}
  \city{Champaign}
  \state{Illinois}
  \country{USA}
}

\author{Lu Ming}
\affiliation{
  \institution{Gallaudet University}
  \city{Washington}
  \state{District of Columbia}
  \country{USA}
}

\author{Sarah Masud}
\affiliation{
  \institution{University of Illinois Urbana-Champaign}
  \city{Champaign}
  \state{Illinois}
  \country{USA}
}

\author{Qi Wang}
\email{qi.wang@gallaudet.edu}
\affiliation{
  \institution{Gallaudet University}
  \city{Washington}
  \state{District of Columbia}
  \country{USA}
}

\author{Yun Huang}
\email{yunhuang@illinois.edu}
\affiliation{
  \institution{School of Information Sciences, University of Illinois Urbana-Champaign}
  \city{Champaign}
  \state{Illinois}
  \country{USA}
}

\begin{abstract}
 Deaf and Hard-of-Hearing (DHH) learners face unique challenges in video-based learning due to the complex interplay between visual and auditory information in videos. Traditional approaches to making embedded audio in video content accessible to DHH learners rely on captioning, but such solutions often neglect the cognitive demands of processing both visual and textual information simultaneously for DHH learners. This paper introduces a set of \textit{motion-driven} design ideas, aimed at support cognitive processing and improving video learning experiences for DHH learners. Through a formative study, we identified four key challenges, including misaligned content and visual overload, and proposed four design suggestions accordingly. Then, we conducted user studies with 16 DHH participants that found improving visual-audio relevance significantly enhance the learning experience by reducing physical and mental demand, alleviating temporal pressure, and improving learning satisfaction. We also found that guiding visual attention and centralizing essential text significantly improved learning satisfaction. However, further research is needed on pacing design; for example, pausing at key moments may yield different effects than evenly slowing down. Our findings highlight the potential of motion-driven design to transform educational content for DHH learners, and we discuss implications for more accessible video-based learning.
\end{abstract}


\maketitle
\section{Introduction}



Efforts to make video content accessible for Deaf and Hard of Hearing (DHH) learners often focus on closed captions and transcripts, for example, \cite{bhavya2022exploring}. 
However, simply providing accurate audio-to-text features does not fully address accessibility challenges \cite{marschark2005access}. Captions alone are insufficient, as they do not account for the visual aspects of lecture videos, which are crucial for comprehension and learning outcomes \cite{bhavya2022exploring, chen2024towards}. DHH learners, on average, read more slowly than their hearing peers, and their constant attention switches between reading captions and watching videos, making it difficult to keep up with content delivery \cite{wang2014we, kushalnagar2010multiple, kushalnagar2014accessibility}. Limited studies in previous work highlight the value of sign language comments in video-based learning, showing how they support peer interaction and provide visual explanations that improve understanding \cite{chen2024towards, chen2024comparison}. Such studies underscore the need for further research on accessible video that goes beyond captioning and centers on enhancing DHH learners' visual experiences and preferences.


DHH individuals process visual information differently from their hearing counterparts, which is crucial for designing effective educational videos to meet their unique information processing needs. Studies show that they respond more quickly to peripheral stimuli, focusing more attention on the periphery \cite{hong1991central,chen2006effects}. Proksch and Bavelier \cite{proksch2002changes} found that while hearing individuals are more distracted by central stimuli, DHH individuals are more susceptible to peripheral distractions, suggesting different allocation of attentional resources \cite{dye2008visual}. Furthermore, DHH individuals demonstrate greater activation in motion-selective visual areas and even the auditory cortex when viewing visual motion \cite{fine2005comparing, bavelier2001impact}, responding faster and more accurately to changes in motion direction \cite{hauthal2013visual}. These visual processing differences should be considered in video design for this target population. DHH individuals' preferences may contribute new insights into previous studies that found that movement, when used strategically, can make information easier to process and interact with on websites \cite{petersen2002eye}.

Our study builds on evidence that DHH individuals exhibit distinct visual and motion processing patterns, which informs how visual content should be designed in educational videos. We introduce the idea of \textit{motion-driven design} to better align video movement with these perceptual preferences, using the term motion to underscore DHH individuals’ heightened sensitivity to visual motion—reflected in greater activation of motion-selective and auditory cortical areas, as well as faster, more accurate responses to motion changes \cite{fine2005comparing,bavelier2001impact,hauthal2013visual}. Motion design in education, such as animations, is often grounded by Multimedia Learning Theory \cite{mayer2002animation}. The theory posits that learning is more effective when information is presented through multiple channels (e.g., visual and auditory) \cite{mayer2009multimedia}. For example, in a science video where a narrator says, ``now the reaction becomes unstable,'' hearing learners rely on tone and timing to guide attention. DHH learners, who rely less on audio and more of captions, may benefit from motion cues—such as the pulsing or shaking of a chemical container—that visually convey the urgency and timing typically communicated through speech.

This paper addresses \textit{How can video presentation designs be improved to better support DHH learners?} We conducted a two-phase study. Phase I, a formative study, answers the research question (\textbf{RQ1}): \textit{``What challenges in the delivery of video lectures hinder the learning experience of DHH learners?''}
We recruited DHH learners and instructors who have experience in deaf education to identify video lecture delivery challenges when watching educational videos and propose improvements. Based on participants' unique perspectives and experiences, we derived four key video lecture delivery challenges that hinder the learning process for DHH learners. For each challenge, we proposed motion-driven design ideas that focus on improving the motion aspect of the video representation. 
Phase II answers the research question (\textbf{RQ2}): \textit{``How do DHH learners perceive the value of the proposed motion-driven design ideas for video presentation?''}
We specifically asked the participants to compare the edited video clips based on the new design suggestions with the original versions. The results showed that two of the design suggestions, that is, making the visuals more relevant and guiding the visual attention, are more effective and highly valued by the participants. 

This research contributes to the HCI, learning science, and accessible computing communities by examining Multimedia Learning Theory in the context of DHH learners—a population that remains underrepresented in this line of research. We identified key barriers in multimedia lecture delivery when learners are primarily using the visual channel, including temporal misalignment between visuals and audio/captions, irrelevant visual content, lack of visual attention guidance, and text overload. In response, we introduce motion-driven design, a set of design ideas developed with DHH learners to address these challenges. Our empirical findings show that these design ideas improved learning satisfaction and supported various cognitive processes in multimedia learning theory, such as managing essential processing and fostering generative processing. While this work defines desired design outcomes for educational videos, improving upstream video production processes remains an area for future research.







\section{Related Works}
In this section, we provide relevant literature review focusing on the visual information processing needs and unique challenges of DHH learners when consuming learning videos and other multimedia content.  Their distinctive visual abilities and challenges as well as the current state of caption call for improved design in learning videos that not only enhance  accessibility but also mitigate visual split-attention, reduce cognitive overload, and promote DHH learners' learning with multimedia content.   

\subsection{Challenges in Captioning and Visual Attention for DHH Learners}

Efforts to make video learning content accessible to DHH learners often focus on closed captions and descriptive transcripts. However, it is a misguided assumption that audio-to-text features alone address accessibility challenges. A key issue is the inaccuracy of captions generated by automatic speech recognition algorithms \cite{bhavya2022exploring}, which are still prone to errors due to noise and ambiguity in speech \cite{kafle2016effect}. Even high-quality captions, such as those on television, have been found to contain errors and readability issues \cite{arroyo2024users}. As a result, autogenerated captions are often too inaccurate to support DHH learners when used alone \cite{parton2016video}. Furthermore, ASR errors are harder to follow than those made by humans in collaborative captioning \cite{kushalnagar2014accessibility}. When asked about future tools, DHH learners expressed interest in improving nonspeech elements such as speaker identity, speech rate, and volume \cite{mcdonnell2021social}.

The appearance of the caption also presents challenges. Unlike standardized TV captioning, most online captions lack defined norms and still reflect outdated standards from the 1970s \cite{kushalnagar2013captions}. Social media platforms that could host educational videos, such as TikTok, also lack consistency, leading to calls for user-generated standards that address elements such as typeface, color, rate, and speaker ID \cite{mcdonnell2024caption}. Additional problems arise when the verbal content and text on the screen present separate information \cite{lasecki2014helping}. Many learners, especially DHH individuals, read more slowly than they listen \cite{jensema1998viewer, tyler2009effect}, and DHH individuals also expend more visual effort tracking captions across multiple speakers \cite{amin2023speaking, amin2023understanding}. As a result, they often miss key information due to visual separation of (split) attention between captions and slide text, a challenge not experienced by their hearing peers. This well-documented issue underscores the need for accessible learning videos that reduce split attention demands for DHH learners \cite{mather2012issue}.

\subsection{Visual Abilities of DHH vs Hearing Learners}
Another important consideration is that DHH learners often demonstrate differences in knowledge structures, conceptual organization, and cognitive strategies compared to their hearing peers, which can place them at a disadvantage in classrooms not designed for this variability \cite{marschark2008cognitive}. Research also shows differences in visual abilities: DHH individuals respond more quickly to peripheral stimuli \cite{hong1991central, chen2006effects}, suggesting enhanced peripheral attention. In a comparative task, DHH participants were more distracted by peripheral distractors, while hearing participants were more affected by central distractors \cite{proksch2002changes}. Rather than increasing distractibility, this may reflect adaptive attention shifts due to reduced peripheral auditory input \cite{dye2008visual}. These perceptual differences are relevant not only for learning outcomes \cite{hauser2008we}, but also for classroom and content design. For example, visually supportive classroom layouts include square-shaped rooms, limited group sizes, and semicircle seating to enhance visual contact \cite{mather2012issue}.

Studies also find differences in motion perception: DHH individuals recruit auditory cortex when viewing visual motion \cite{fine2005comparing} and show increased activation in motion-selective visual areas \cite{bavelier2001impact}. They also outperform their hearing peers in detecting small directional changes in motion \cite{hauthal2013visual}. These findings underscore the need for video content to account for the increased peripheral and motion sensitivity. Although hearing learners may tolerate complex visual scenes, DHH learners may find them distracting or cognitively demanding. This is especially relevant in mobile learning contexts where environmental distractions are greater than in static settings such as lectures \cite{jain2018towards}.

\subsection{Improving Multimedia Learning Experience for DHH Learners}

Multimedia represents concepts through text, images, video, sound, and animation. Although beneficial, poorly designed multimedia can cause cognitive overload – when task demands exceed the learner’s processing capacity \cite{mayer2003nine}. According to Multimedia Learning Theory, humans process information through dual channels (visual and auditory/verbal), each with limited capacity, and active learning requires selecting, organizing, and integrating content \cite{mayer20143}. More recent work expands these principles with refined evidence-based design strategies \cite{mayer2024past}, including signaling (highlighting key content), coherence (removing extraneous material), and segmenting (breaking content into manageable parts) \cite{mayer2021evidence}. Effective multimedia minimizes the strain on each processing pathway to support three types of cognitive processing: essential, extraneous, and generative.

Motion design in educational videos supports multimedia learning by guiding attention, clarifying complex concepts, and maintaining engagement through dynamic visuals such as animated annotations, synchronized highlights, and pacing adjustments \cite{mayer2003cognitive, boucheix2010eye}. Although principles like \textit{ temporal contiguity} and \textit{signaling} are well-established for hearing learners \cite{mayer2009multimedia, sweller2011cognitive}, they are rarely adapted to the unique visual processing strengths of DHH individuals or account for the absence of an audio channel. Our study expands this foundation by exploring motion design strategies that better align with DHH learners’ perceptual preferences and cognitive demands \cite{chen2025motion}.

Information and communication technologies can improve learning for DHH students when tailored to their needs \cite{debevc2004role}, but design guidance remains limited. Existing theories often assume an auditory/verbal channel, which is absent for DHH learners \cite{techaraungrong2017design}. Few studies focus on reducing cognitive overload in accessible multimedia for DHH learners \cite{hidayat2017multimedia}. For example, pairing font color with font weight showed low cognitive load in affective captioning \cite{de2024caption}, yet preferences varied widely. Similarly, preferences for interpreter placement and font size differ between individuals, highlighting the need for customizability \cite{boudreault2024closed}.

Accessible multimedia—such as captions or sign language—can still create cognitive overload, contributing to DHH learners gaining less from lectures than hearing peers \cite{lang2002higher}. While some learners benefit from onscreen instructor images, others find them distracting \cite{kizilcec2015instructor}. Gu et al. \cite{gu2024onscreen} found that social presence is only helpful when its cognitive benefit outweighs the attentional cost, consistent with the seductive details principle \cite{mayer2020five}. Studies also show that DHH learners often focus on text and overlook visualizations when both are presented \cite{schmidt2010closer}. Unlike hearing learners who divide attention across modalities, DHH learners must shift attention between multiple visual sources, such as captions and slides \cite{marschark2005access}.


\section{Formative Study: Identify Video Lecture Delivery Challenges and Propose Design Ideas (RQ1)}
In Phase 1, we understood the challenges in video lecture delivery for DHH learners to consume videos created for a general audience, focusing on video learning as a multimedia learning experience rather than limited to caption design. 
In our manuscript, ``mainstream video'' refers to video created for a general audience. The term ``mainstream'' is used analogously to its use in educational contexts, where it denotes the practice of integrating learners with special education needs into general education classrooms based on their abilities \cite{lindsay2007educational}.  The research team identified four key challenges associated with making mainstream videos accessible to DHH learners and proposed four corresponding motion-driven design ideas(challenge-suggestion pair). 
Furthermore, the research team and most participants agreed that the term ``video learning challenge'' was inadequate, as it placed emphasis on the process of learning of DHH learners rather than the shortcomings of the video material. Therefore, we prefer the term ``video lecture delivery challenges'' to more accurately reflect the source of the difficulties.

\subsection{Study Details and Procedure}

\subsubsection{Positionality}
The research team consisted of researchers with diverse hearing abilities and backgrounds. The first author has hearing with beginner-level ASL skills and led this phase of the study. A Deaf co-author and three hearing co-authors also participated in this phase. The study was supervised by two hearing faculty members, one of whom has more than 30 years of experience in college-level DHH education and is fluent in ASL. Data collected in ASL were transcribed into English for team discussions by the Deaf coauthor and a hearing faculty member with fluent in American Sign Language (ASL).

\subsubsection{Study Overview}

Six DHH college students (five self-identifying as Deaf and one as Hard-of-Hearing) and three instructors with experience in DHH education (one Deaf and two hearing) participated in this phase. The student participants majored in design, business, education, or computer science and self-reported that they had some understanding of AR and could explain key concepts. Four of the nine participants identify as male and five as female. The collection of video lecture delivery challenges and suggestions was conducted asynchronously, with participants completing a provided canvas. The participants spent between two and four hours in total on this task. Afterwards, they met synchronously with the research team to clarify any information noted on the canvas. This approach was chosen because fully annotated sessions were perceived as too time consuming for synchronous interaction. Each student participant was compensated \$25 per hour for their time, while the three instructors voluntarily participated in the study. We tracked the emergence of new codes after each participant and nine participants were recruited until saturation of the design suggestion themes was achieved. We consider this phase as a formative study that prioritizes the participation of DHH learners and individuals with lived experiences. 
Motion-driven design subject matter experts were not included as participants because no individuals with knowledge of DHH individuals were identified during initial searches. Additionally, accessibility is rarely mentioned in the existing literature on motion design education, suggesting that accessibility is unlikely to have been taught to motion designers. 

During the study, participants were asked to watch an educational video on Augmented Reality (AR) technology created by Coursera for mainstream learners, identify challenges in video lecture delivery, and describe the corresponding changes they wish to make. 
This mainstream educational video was the first lesson in a series of courses on AR technology. It covered the history of AR, widely known applications of AR, and the major technical areas of AR. The video presentation style alternated between a talking head\footnote{A talking head video is a video of someone speaking directly in front of the camera. It is shot in a way that the viewer feels that the speaker is talking to them face to face. talking-head is a common video lecture production style \cite{guo2014video}.} and full-screen visual examples and is sometimes included both simultaneously. We chose this video for our study as an example of a mainstream video developed for hearing learners with only spoken English used by the instructors but no ASL interpreters present. The video was selected because it had the most reviews (average score 4.5 out of 5 by 3.7k learners) on Coursera on the AR topic, and was marked beginner-level. In other words, it was perceived as helpful and well designed as a mainstream video. The research team added error-free open captions to the video. We decided to include only one video for the study to manage the overall duration of the study session based on the availability of our participants, and we acknowledge that this single video may not fully explore the challenge of video lecture delivery that exists in all educational videos.

\subsubsection{Step 1: Identifying Video Lecture Delivery Challenges and Suggestions for DHH Learners}

First, participants were asked to identify the challenges in video lecture delivery and suggestions for the 15-minute mainstream video about AR technology. The researcher explained to the participants that this video was an example of a mainstream video and that the participants were encouraged to recall their previous watching experience of mainstream videos to focus on challenges that have been experienced frequently before. 
Each participant was given a canvas to list challenges and suggestions related to various video clips or screens within the video. The canvas contained multiple rows, with each row representing a challenge–suggestion pair tied to a specific video timestamp in google spreadsheet (see example in Fig.~\ref{fig:sample1}). Participants were asked to provide as many challenges-suggestions pairs as they could.  Multiple occasions of the same challenge-suggestion pair would be listed as one. For example, if the clips 2:10–2:40 and 3:40–2:55 share the same challenge-suggestions pair, they should be combined and listed as a single entry.

Participants were encouraged to use approaches they felt helpful to adequately express challenges and suggestions, such as typing text, drawing out their ideas, or having face-to-face interactions with the research team. Previous research highlights the importance of supporting the preferred modalities of DHH individuals in research participation, as English is often not their preferred language \cite{chen2023exploring}. Annotated examples are shown in the appendix \ref{appendix:designexamples}.
In total, we collected 105 challenge-suggestions pairs from our nine participants. 

\textit{Additional Material} We had some initial challenges explaining the purpose of our study to participants, as many were hesitant to share ideas without knowing who would actually implement the proposed changes. Some participants intentionally suggested fixes they believed would be easier for instructors to adopt. To encourage more open thinking, the research team showed short videos demonstrating how generative AI could easily modify video content, helping to reduce some of these concerns and making the design process feel more attainable.




\subsubsection{Step 2: Explaining How Design Suggestions Address DHH Learners' Needs}
Next, participants were asked to explain how each challenge-suggestion pairs would improve learning the video for DHH learners by selecting checkboxes on the canvas. The checkboxes were used mainly to help the participants understand the purpose of our study: empowering the visual abilities of DHH learners and supporting multimedia learning. Participants can revise their responses to Step 1 if they wanted as they gain a better understanding of the purpose of the study.  

There are a total of seven checkboxes. We explain each concept to participants prior to each study using both written English and ASL. The first three checkboxes included the following visual abilities that have been studied in \cite{bell2019cross}: 
\textit{Spatial Vision}: Understanding where things are, how they are arranged and how they look;
\textit{Temporal Vision}: The ability to perceive and process visual information over time;
\textit{Motion Vision}: Noticing and understanding movement, compiling both spatial and temporal information. 
The other four checkboxes included addressing the following cognitive demands according to the Multimedia Learning Theory \cite{mayer2002multimedia}: 
\textit{Reducing Irrelevant Information};
\textit{Focusing on Essential Information};
\textit{Fostering Connection between Text and Image} and ``Others,'' which can be selected if none of the three demands can be applied. The three checkboxes map to 3 types of cognitive demands according to Multimedia Learning Theory as explained in related works:  Essential processing, Extraneous processing, and 
Generative processing. Our checkboxes are reworded versions of the three demands, revised based on the reading preferences of DHH learners in consultation with members of the Deaf community. The checkboxes were not used as a data collection method; instead, they served as study materials to convey the purpose of the study and ensure that the intended purpose was communicated effectively. The process was particularly helpful in clarifying how specific terms should be explained in preparation for the later study, as many scientific terms created in English do not have an equivalent ASL term.

\subsubsection{Analysis} 
\label{analysis}
The research team conducted a thematic analysis \cite{braun2012thematic} of the challenges and suggestions for video lecture delivery, iteratively refining the themes through weekly meetings and discussions. The analysis began with data collected from the first six participants, and additional participants were recruited and included in the data analysis one by one until no new themes emerged. To facilitate discussions, a team member segmented the 15-minute video into 17 clips based on natural pauses in the content. Each clip lasted between 30–120 seconds (M = 53.4, SD = 23.27) and received 5 to 10 challenge-suggestions pairs based on the timestamps provided by the participants. 

The first and last authors initiated the analysis by independently reviewing participant data and tentatively proposing eight initial themes based on responses from the first six participants using inductively. The sample themes included in ``visual and audio/captions are not temporally aligned: increase alignment,'' ``too much visual information at the same time: reduce visuals,'' and ``visual content irrelevant to audio / captions: remove visuals.'' During weekly meetings, the team reviewed these themes to assess their sufficiency in capturing patterns within the data. For each video clip, team members selected the themes they found relevant, and the team counted how many video clips received the same selected codes from all team members as an indicator of agreement. The initial agreement rate among team members was 70.6\% (12/17 clips). Through iterative discussions, the themes were refined and expanded to ensure they comprehensively covered the data and agreement among team members. Disagreements were resolved by revisiting the raw data, comparing themes, and merging similar themes, leading to the identification of four final themes that cover more than 80\% of the initial 105 pairs.

The weekly discussions focused on refining the themes to ensure that they captured participants' data comprehensively, rather than merely determining and categorizing which theme best fits a specific video clip. Theme definitions were updated after each weekly discussion. Table \ref{tab:RQ1findings} (columns 1–3) summarizes the final design ideas (challenge-suggest pair), challenges in video delivery, and suggestions provided by participants derived from thematic analysis. The table serves as a summary of themes derived from thematic analysis and is not merely a classification of participant suggestions.




\begin{table}[hb!]
    \centering
    \renewcommand{\arraystretch}{1.5}
    \begin{tabular}{ p{1.6cm}p{3.2cm}p{3.5cm}p{4.5cm}}
      \textbf{Design Idea}   & \textbf{Challenges in Video Lecture Delivery} & \textbf{Suggestions} & \textbf{Changes to Make (by video lecture creator)} \\
      \hline
      \hline
       \textit{D-Illustrate}: Improve Visual-Audio Relevance.  &Lack relevant visual elements to explain the captions. & Replace such visuals with illustrative elements for the captions and static ``talking-heads'' \cite{guo2014video}. The new content should not be more visually demanding than the old one.  &Replace old visuals with new ones that illustrate the captions to reduce the gap between the semantic meaning of the visuals and the audio. No change to the audio.
       \\ 
       \textit{D-Align}: Align Visual Attention & Important visual elements are shown all at once or out of sync with the audio/captions, causing split attention, confusion, and difficulty to comprehend. & Overlay shades, shapes, colors, and animation effects to guide visual attention, and synchronize visual elements with the corresponding audio or captions. & Identify key visual elements and when they are referred in the audio. Add sequential visual cues and adjust visual timing to match audio or captions. Alternatively, insert brief pauses or extend visual display time for clarity and pacing.\\

       \textit{D-Declutter}: Centralize Essential Text  & Lack visual support: screens jammed with captions filled with repeated text chunks, increasing unnecessary reading load and causing anxiety.  & Visually center and summarize the text block to make the key information stand out. Remove/reduce lengthy irrelevant visual information including static ``talking head.''  & Identify long and duplicated texts on the screen and in the captions, extract essential information/terms with visual meanings, re-design the visual typographic of captions to convey the meaning visually.  No change to the audio.  \\
       \textit{D-Slowdown}: Slow Visuals Down  & Lack time to comprehend video content: rapid visual movements hinder learners from reading captions and causing anxiety and distraction. &Slow the visual movements and extend the on-screen time so there is enough time to view/read.  &Extract fast-moving visuals and extend their play time on screen. Add audio pauses if needed. Visual and audio may both be relocated, and video might be longer. \\
    \end{tabular}
    \caption{RQ1 findings: Motion-driven Design Ideas.  
    }
\label{tab:RQ1findings}
\Description{This table summarizes the design Ideas, challenges in video lecture delivery they addressed, suggestions, and changes to make by video lecture creator. For ``D-Illustrate: Improve Visual-Audio Relevance,'' the challenges are ``Lack relevant visual elements to explain the captions;'' the suggestions are ``Replace such visuals with illustrative visuals for the captions, including a 'talking-head'. The new content should not be more visually demanding than the old one;'' and the changes to make are ``Replace old visuals with new ones that illustrate the captions to reduce the gap between the semantic meaning of the visuals and the audio. No change to the audio.'' For ``D-align: Align Visual
Attention
and Timing'', the challenges are ``Important visual ele-
ments are shown all
at once or out of sync
with the audio/captions,
causing split attention,
confusion, and difficulty
following along.;'' the suggestions are ``Overlay shades, shapes, col-
ors, and animation effects
to guide visual attention,
and synchronize the ap-
pearance of visual elements
with their mention in audio
or captions;'' and the changes to make are ``Identify key visual elements and
when they are referenced in the au-
dio. Add sequential visual cues and
adjust visual timing to match audio
or captions. Optionally insert brief
pauses or extend visual display time
for clarity and pacing..'' For ``D-Declutter: Centralize Essential Text,'' the challenges are ``Lack visual support: screen filled with chunks of repeated text in captions, increasing unnecessary reading load and causing anxiety;'' the suggestions are ``Visually center and summarize the text block to make the key information stand out. Remove/reduce lengthy irrelevant visual information accordingly, such as 'talking head';'' and the changes to make are `` Identify long and duplicated texts on the screen and in the captions, extract essential information/terms with visual meanings, re-design the visual typographic of captions to convey the meaning visually.  No change to the audio.'' For ``D-Slowdown: Slow Visuals Down,'' the challenges are ``Lack time to comprehend the video content: rapid visual movements hinder learners from reading captions and causing anxiety and distraction;'' the suggestions are ``Slow the visual movements and extend the on-screen time so there is enough time to see/read;'' and the changes to make are ``Extract fast-moving visuals and extend their play time on screen. Add audio pauses if needed. Visual and audio may both be relocated, and video might be longer.''}
\end{table}

\subsection{Finding: Four Video Lecture Delivery Challenges and Suggestions}

The motion-driven design is for delivering multimedia content more efficiently along the timeline. Four motion-driven design ideas were proposed based on the challenges in lecture delivery identified by our participants: \textit{D-Illustrate}, \textit{D-Align}, \textit{D-Declutter}, and \textit{D-Slowdown}. \textit{D-Illustrate} addressed the lack of relevant visual elements to explain captions by replacing such visual elements with illustrative visuals while not making it more visually demanding. \textit{D-Align} addressed the lack of focus on visual elements by adding overlay shades, shapes, colors, and animation effects to effectively guide viewers' attention. \textit{D-Declutter} addressed the lack of visual support that increased unnecessary reading load by visually centering and presenting the text block to make the key information stand out while removing or reducing irrelevant visuals. \textit{D-Slowdown} addressed the lack of time to fully comprehend video content by slowing visual movements and extending the time on the screen to allow enough time to see/read.
The four video lecture challenges are presented in Table \ref{tab:RQ1findings} column 2. For each of the four identified video lecture delivery challenges, the corresponding motion-driven design suggestions are presented in columns 3 of Table \ref{tab:RQ1findings}.  Our participants did not suggest changes to the audio or caption content as they identified it as important lecture material. 
Note that this ranking does not reflect the importance of the design ideas themselves, nor does it indicate their significance. Since a pair of suggestions collected from participants may be associated with more than one of the four design ideas, we did not provide the exact number of pairs to reduce confusion. 

\begin{figure}[ht]
    \centering  
    \includegraphics[width=1.0\textwidth]{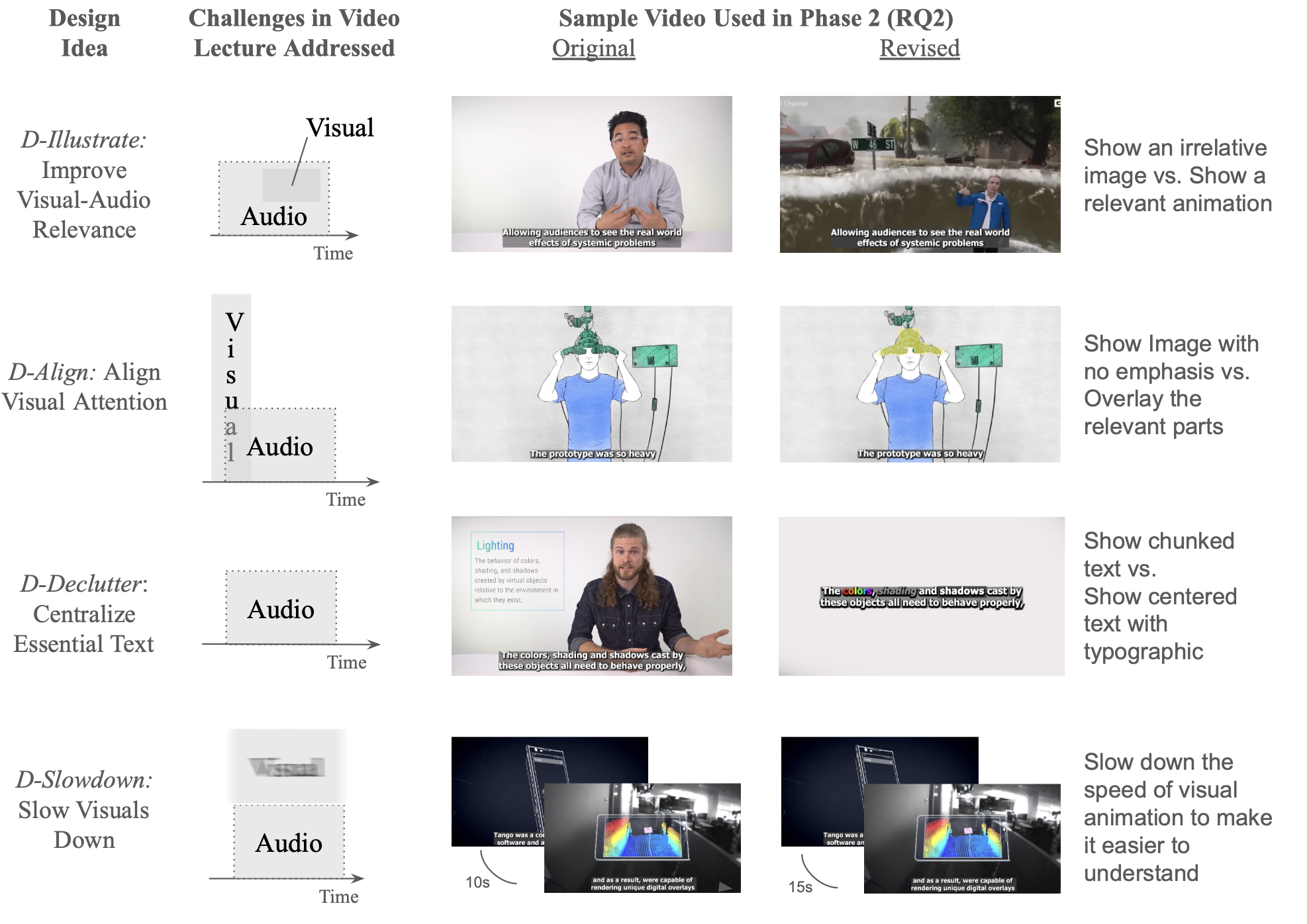}    
    \caption{Original vs Revised Video Pairs used in RQ2 Following Four Motion-Driven Design Ideas (challenge-suggestion pairs) Identified in RQ1.
    } 
    \Description{This figure describes the four video examples used in RQ2: original vs revised. There are three columns in this image: design idea, challenges in video lecture addressed, and sample video used in phase 2 (RQ2). The sample video column has three sub columns: Original, Revised, and annotation of the differences. For D-Illustrate: improve visual-audio relevance, the challenge is visual only shown in part of the audio. The original screenshot is a man talking, whereas the revised screenshot is showing some animation. The annotation is: show an irrelevant image vs show a relevant animation. For D-Align: align visual attention, the challenge is visuals are not aligned with the audio. The original video shows a man using an AR device, whereas the revised video highlights the device in yellow. The annotation is: show image with no emphasis vs. overlay the relevant parts. For D-Declutter: centralize essential text, the challenge is there is only audio on the timeline. The original video shows a man talking with captions below and text illustration beside. The revised video shows only caption with color in the middle of the screen. The annotation is: show chucked text vs. show centered text with typographic. For D-Slowdown: slow visuals down, the problem is the visual moves too quick. The original video took 10 seconds between two screenshots, whereas the revised took 15 seconds. The annotation is: slow down the speed of visual animation to make it easier to understand.
}
    \label{fig:screenshot}
\end{figure}

\subsection{Implication and Implementation for RQ2 Study} \label{sec:rq2prep} 

Following columns 1-3 of Table \ref{tab:RQ1findings}, the research team revised the same 17 clips from the 15-minute video accordingly. The changes were refined through multiple conversations with DHH members who participated in the initial video annotation phases. Column 4 in Table \ref{tab:RQ1findings} summarizes all key changes that the creators of the video lecturer made. It does not include fine-tuned feedback on specific details for brevity, such as adjusting playback speed (e.g., 2x vs. 3x) or color preferences (e.g., yellow vs. red). 

The decision of which of the four design ideas to apply to which video was agreed by the research team during the analysis phase described in section \ref{analysis}. There are six video clip pairs for \textit{D-Illustrate}, five video clip pairs for \textit{D-Align}, three video clip pair for \textit{D-Declutter}, and three video clip pairs for \textit{D-Slowdown}. Examples are shown in Fig. \ref{fig:screenshot}. The number of clips under each design idea largely depends on the challenges present in the original video, so we focused on addressing the challenges in each clip rather than attempting to evenly distribute the number of clips for each design idea. 
Additionally, we do not claim that our design ideas sufficiently summarized all challenges in motion-driven design or indicated whether one video lecture delivery challenge was more common than another. These are problems that require further research. It is important to note that the challenge-suggestion pairs are directly linked. The choice of suggestion entirely depends on the specific nature of the original challenge being addressed. 
 
For RQ2 study design purposes, each clip was applied with one single design idea. Based on our findings, three of the 17 clips appeared to have more than one challenge-suggestion pair. The researchers chose to focus on the most significant challenge for these clips. To minimize the impact of secondary challenges, the researchers first addressed the secondary challenge of creating a revised ``original'' version of each video. Then, they made further modifications to address the main challenge for comparison in the study. The research team recreated these three video clips to closely resemble the original online video, although some slight differences remained, such as font and color shading. In summary, 14 of 17 ``original'' video clips used in the study were the exact versions available online, while the other three video clips included some manual edits for comparison purposes of the study design. When we refer to the ``original'' version in RQ2, we refer to the video with only one video lecture delivery challenge that was improved based on.

The revision of the original video was done manually using professional video editing software.  We explored the feasibility of using AI tools to make these changes, but found them insufficient, leading to the decision to proceed with manual edits. More reflections on the try-out AI tools are presented in the discussion section. For each video clip, two editors that are proficient users of Adobe PR to create multimedia learning material worked together. The first editor reviewed the clip and, referred to Table \ref{tab:RQ1findings}, based on the input of the participants previously collected and discussed with the research team, proposed changes by drawing and annotating key frames. After completing their edits independently, the second editor reviewed the proposed changes to ensure they aligned with the intended design directions and requested revisions from the first editor when necessary. The finalized version of each edited video clip was reviewed by the research team to ensure consistency and alignment with the ideas proposed. Two pilot testing was done to refine the editing details, such as exact speed of slowing in D-slowdown.


\section{Evaluating Perceived Value of Design Ideas (RQ2)}

\subsection{Method}

In this phase of the study, we conducted user studies with DHH learners  (n=16) to review the perceived value of four design ideas identified in the previous formative study. The participants were referred to as V1 to V16 in the remainder of this paper.
Each study session took between 1.5 and 2 hours. We compensated the participants at a rate of \$25 per hour. This study is approved by the Institutional Review Board (IRB).

\begin{figure}[ht]
    \centering  
    \includegraphics[width=1.0\textwidth]{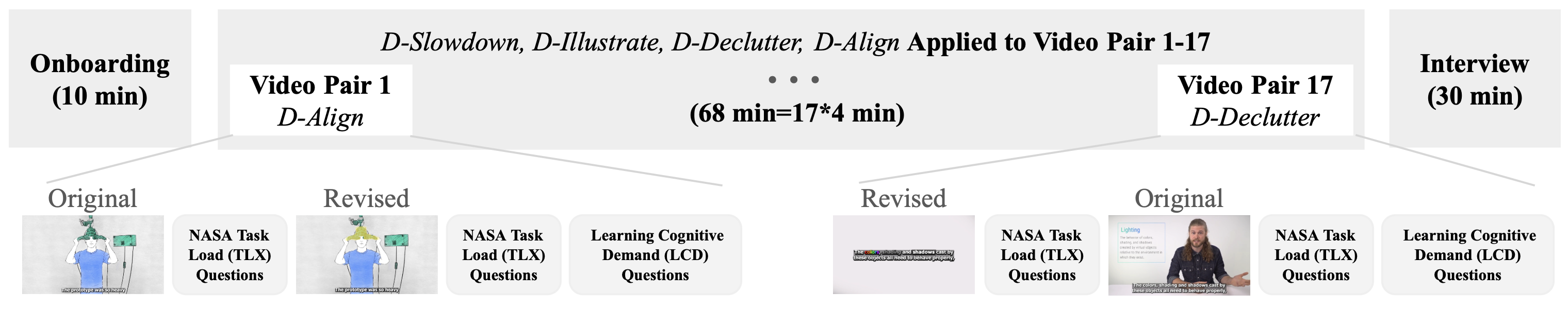}
    \caption{Study Procedure for RQ2. During onboarding, participants were introduced to all design ideas described in Table \ref{tab:RQ1findings}. Then, they watched 17 video pairs. Each pair of videos consists of an original (unedited) version and a revised (edited) version with one of the four design ideas applied. For each pair of videos, participants will watch both versions in a randomized order (either showing the original version first, or the revised version first, as demonstrated by two examples in the figure), and then complete a four-question survey developed from NASA Task Load Index for each version they watched (TLX questions). Then, they will answer three additional survey questions on learning cognitive demand scores (LCD questions) comparing the two versions. The same survey questions were used for all 17 video pairs. Participants were given at least one 5-minute break during video rating. Then, participants took place in a brief interview to discuss about their suggestions toward the design ideas they experienced. 
    } 
    \Description{This figure introduces the study process. It is built with two main lines. In the first line, there are three main grey boxes showing three steps of the study procedure. The first box is “Onboarding (10 minutes).” On the top of the second box, it writes D-Slowdown, D-Illustrate, D-Declutter, and D-Align Applied to Video Pair 1 to 17 (68 minutes = 17*4 minutes). There are two white boxes representing Video Pair 1 (D-Align) and Video Pair 17 (D-Declutter) on two sides of this box, with ... in between. The white box for Video Pair 1 is expanded in the second line with the following: screenshot of the original video, text "NASA Task Load (TLX) Questions," screenshot of revised video, text "NASA Task Load (TLX) Questions," text "Learning Cognitive Demand (LCD) Questions." The white box for Video Pair 17 is expanded in the second line with the following: screenshot of the revised video, text "NASA Task Load (TLX) Questions," screenshot of original video, text "NASA Task Load (TLX) Questions," text "Learning Cognitive Demand (LCD) Questions." The third box is “interview (30 minutes).”
}
    \label{fig:procedure}
\end{figure}

\subsubsection{Positionality}
The same research members in formative study worked on the current phase. The first author designed, conducted, and analyzed the study. A Deaf co-author conducted 12 studies, whereas the remaining studies were led by the first author with ASL interpreters for communication with participants. All study sessions were supervised by the hearing faculty with experience in DHH education. The study data was mainly analyzed by the first author and two other hearing coauthors with the assistance of the Deaf coauthor. 

\subsubsection{Participants}

We recruited 16 DHH participants from a university in the US that specializes in DHH education. None of the participants in this phase participated in the previous phases of the study. The participants were between 20 and 49 years old. Of all participants, 14 self-identified as Deaf, while the other two self-identified as Hard of Hearing. Ten were self-identified as male, while the other six were identified as female. Regarding ethnicity, six self-identified as white, five as African American, three as Asian, one as Hispanic, and one did not disclose. Eleven participants were majoring in business or accounting, four in information technology-related majors, and one in art history. Nine participants reported a sign language (e.g., ASL) as their first language for face-to-face communication, whereas the other seven reported a spoken language (e.g., English) as their first language for face-to-face communication. All participants were recruited through word of mouth and posters posted on the university campus.  Participants in this phase reported having no prior knowledge or experience with AR.


\subsubsection{Study Procedure}

We collect feedback from DHH learners on our proposed design ideas asking participants to compare the video clip pairs (original and revised) developed during formative study using the within-subject approach. Each pair demonstrated one of the design ideas in Table \ref{tab:RQ1findings}, allowing participants to evaluate and provide input on the proposed changes. 
All studies were conducted through Zoom synchronously in the procedure described in Figure \ref{fig:procedure}. After completing the consent form, participants were first introduced to the four design ideas as an onboarding process. The researchers explained the terms used and encouraged participants to ask any questions they might have. 

\textbf{Randomized Original vs. Revised Video Rating.} For each design, we presented two versions, original (before edit) and revised (after edit), in a randomized order. We used the same order to present 17 pairs of video clips based on their presentation order on Coursera, as the original video was designed in a progressive order in presenting knowledge that could lead to confusion if the order was shuffled. We described the four design ideas as ways future technology could be used to improve a mainstream educational video. The participants were told which of the four design ideas was applied to the revised video clip, but were not informed which version was the revised clip. After watching each clip, they were asked to complete the following survey questions on the same page, and modifications to the responses are allowed when on the same page:

\begin{itemize}
    \item \textbf{Learning Cognitive Demands (LCD) Questions}: For each video pair, participants were asked to compare the original and revised versions and rate how the revised version helped participants in three cognitive demands in Multimedia Learning Theory \cite{mayer2002multimedia}: \textit{Reducing Irrelevant Information}, \textit{Focusing on Essential Information}, and \textit{Fostering Connection between Text and Image}. Participants may choose from Strongly Disagree, Disagree, Slightly Disagree, Neutral, Slightly Agree, Agree, Strongly Agree. A total of three questions are completed per video pair. We selected this scale because it is widely used in research on video-based learning in education, which captures more nuanced learning processes than TLX.
    \item \textbf{NASA Task Load Index (TLX) Questions}: For each version of the video clip, participants complete four questions on how each version made participants feel in Mental Demand, Physical Demand, Temporal Pressure, and Learning Satisfaction.  These four questions were developed based on NASA Task Load Index (TLX) questions, which have been used to understand DHH individuals' interaction with technology, for example, \cite{li2022soundvizvr,dust2023understanding, chen2024towards}. Participants could choose on a 10-point Likert scale, where 1 represents low demand / pressure / satisfaction, and 10 represents high demand / pressure / satisfaction. A total of eight questions are completed per video pair--four questions for each version. This scale was chosen due to its widespread use in HCI research to measure task demand \cite{kosch2023survey}. We slightly adjusted it to meet the educational context.

\end{itemize}


To avoid participants from feeling fatigued watching video clips, we included a 5 minute break after watching 11 video clips, when the four design ideas were experienced by participants at least once and around 1 hour into the study session. After the break, the researcher had a brief interview with the participants to review the four design ideas for suggestions and potential additional designs that should be introduced. Then, they continued to watch the remaining 6 video clips. Participants were also allowed to take additional breaks if they wished to during this session. After rating all video clips, they were asked to complete a demographic survey and proceed to the interview. Familiarity with the content of watching the original and revised versions in succession could have influenced the participants' ability to accurately assess cognitive demands. Our study design addresses this challenge as effectively as possible, given the typically limited sample sizes that can be accessed in accessibility studies.

\textbf{Interview.} During the interview, we focused on understanding the feelings of the participants toward the design ideas of the video, as well as their suggestions for further improving the video clips they watched. We also asked participants questions to envision how future technology could automate the video editing process. The sample questions are: \textit{How would you like the video to be further enhanced?}, \textit{Do you think you would directly apply the design ideas to improve accessibility of a video?}, \textit{Do you think all DHH learners would be interested in using the design ideas? Why?}. Participants completed the interview in ASL or spoken English according to their preferences. The interview recordings were transcribed into deitentified transcripts in written English and destroyed immediately thereafter.

\subsubsection{Data Analysis}

For survey questions, we began by examining whether our motion-driven design ideas collectively were perceived as helpful in supporting different aspects related to multimedia learning and task load demands, presented in Section \ref{sec:overview}. Then, we analyze the effect of each design idea. The results are presented in section \ref{sec:findings_lcd} for LCD questions and in section \ref{sec:findings_nasa} for TLX questions. The LCD questions evaluated the perceived helpfulness of each design idea, while the TLX questions provided standardized cognitive load measures that future researchers can refer to. For the LCD questions, we converted the agreement scores (Strongly Disagree, Disagree, Slightly Disagree, Neutral, Slightly Agree, Agree, Strongly Agree) to 1 to 7, respectively.
For the TLX questions, which used a scale of 1-10, each participant rated both the original and revised versions of a video, resulting in a difference score ($\text{Revised} - \text{Original}$) for each question. We then computed estimated marginal means (emmeans) to account for individual differences among participants.

The total number of responses per design idea varies depending on the number of video clips applied to each design idea. For example, in the case of the \textit{mental demand} for the design idea D-Slowdown (which was applied to three clips), each participant provided three ratings for the mental demand of the original video and three ratings for the mental demand of the revised video. Linear mixed models were constructed for both the LCD and TLX questions (using emmeans). We chose linear mixed models as they fit the distribution of our survey data the best and addressed the effects of different numbers of video clips for different design ideas in our study, resulting in differences in the number of responses per participant. For all linear mixed models, we included a random effect for Participant ID (``1/PID'') to account for individual differences that were not explained by the fixed effects in the model \cite{chang2023citesee}. Post-hoc analysis using Satterthwaite's t-tests were conducted to compare the effects of different design ideas. All statistically significant results reported had a power of at least 0.80 and remained significant after applying the Bonferroni correction. The Bonferroni correction is a conservative method for controlling false positives in multiple comparisons by making the significance threshold stricter.

For interview transcripts, we performed an inductive thematic analysis \cite{braun2006using}. Two hearing co-authors independently open-coded two different transcripts and developed an initial codebook from scratch to better understand the data. The initial coding centered on participants’ feedback for each design idea, followed by discussions on how this feedback aligned with—or diverged from—the modeling results. The co-authors then iteratively refined the codebook while collaboratively analyzing additional transcripts. Throughout the analysis process, the hearing co-authors regularly consulted with the Deaf co-author, who led the study, as well as hearing faculty members with expertise in DHH education.  Detailed results are presented in section \ref{sec:findings_interview}.

\subsection{Finding-Perceived Value of Design Ideas} 

\subsubsection{Motion-Driven Design Supports Multimedia Learning and Decreases Task load} \label{sec:overview}


\begin{figure}[t]
    \centering  
    \includegraphics[width=1.0\textwidth]{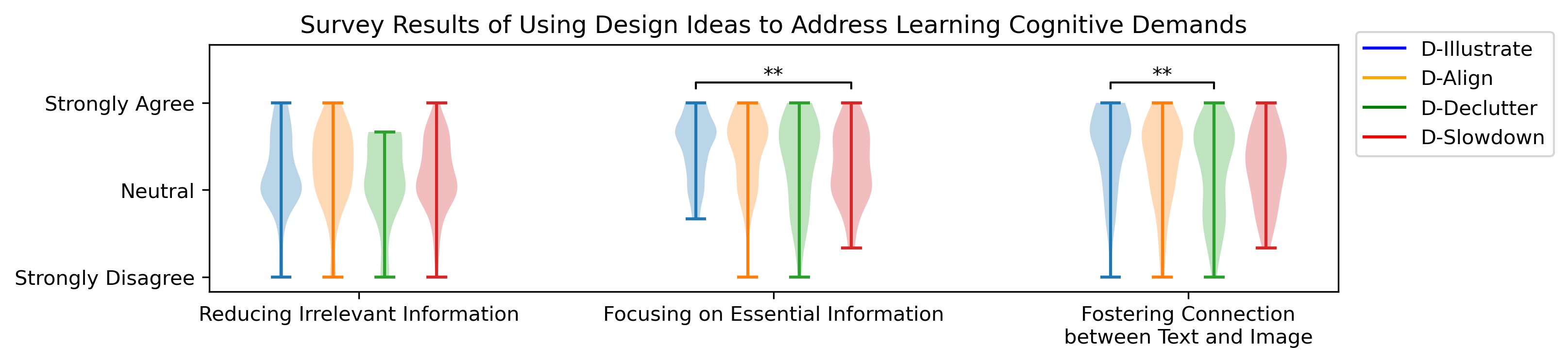}    
    \caption{Survey results of using design ideas to support multimedia learning (LCD Questions), shown in a violin plot. Video clips for each design idea were aggregated. Post-hoc analysis on linear mixed models for each question suggested that for \textit{Focusing on Essential Information}, there is significant difference between D-Illustrate and D-Slowdown. For \textit{Fostering Connection between Text and Image}, there is significant difference between D-Illustrate and D-Declutter. The ** denotes significant differences between the two design ideas at p<.01.} 
    \Description{This figure is a violinplot with title ``Survey Results of Using design ideas to support Learning Cognitive Demands.'' The vertical axis includes Strongly Agree, Neutral, and Strongly Disagree from top to bottom; the horizontal axis includes Reducing Irrelevant Information, Focusing on Necessary Information, and Fostering Connection Between Text and Image. For each label on x-axis, there are four violinplots representing four design ideas evaluated in RQ2: D-Illustrate, D-Align, D-Declutter, D-Slowdown. For Reducing Irrelevant Information, the distribution is similar across four design ideas. For Focusing on Necessary Information, the distribution of D-Illustrate, D-Align, and D-Declutter are skewed toward Strongly Agree, whereas D-Slowdown is evenly distributed. This a line between D-Illustrate and D-Slowdown with ``**'' representing significant difference between the two. For Fostering Connection Between Textand Image, the distribution of D-Illustrate and D-Align are skewed toward Strongly Agree, whereas D-Decluter and D-Slowdown are evenly distributed. This a line between D-Illustrate and D-Declutter with ``**'' representing significant difference between the two.}
    \label{fig:compare_violin}
\end{figure}






Overall, participants rated Slightly Agree on LCD questions across all video clips and question (M = 5.0, SD = 1.42). As shown in Fig. \ref{fig:compare_violin}, participants rated higher agreement for  \textit{Focusing on Essential Information} (M = 5.3, SD = 1.32) and \textit{Fostering Connection between Text and Image} (M = 5.1, SD = 1.47), compared to \textit{Reducing Irrelevant Information} (M = 4.6, SD = 1.40). 
Linear mixed model revealed a significant main effect in different cognitive demands ($F(2, 798) = 18.03$, p < .001). Post-hoc analysis using Satterthwaite's t-tests showed that \textit{Focusing on Essential Information} was rated significantly higher than \textit{Reducing Irrelevant Information} (t(798) = 5.81, p < .001), and \textit{Fostering Connection} also received higher ratings than \textit{Reducing Irrelevant Information} (t(798) = 4.23, p < .001). There was no significant difference in ratings between \textit{Focusing on Essential Information} and \textit{Fostering Connection} was not significant. 
These results indicate that participants found the design more helpful in focusing on essential content and fostering connections, with a lower perceived benefit for reducing irrelevant information.

For TLX questions, linear mixed models and post-hoc analysis suggested that participants rated the revised videos as requiring less cognitive demand and offering more learning satisfaction.
Specifically, there is a significant reduction in mental demand (EMM = –0.51, t(22.3) = -2.56, p < .05), significant reduction for physical demand (EMM = –0.52, t(22.3) = -2.62, p < .05), significant reduction in temporal pressure (EMM = –0.49, t(22.3) = -2.47, p < .05), and significant increase in learning satisfaction (EMM = 0.74, t(22.3) = -3.69, p < .01). Not that the four TLX questions were not compared as each represents a distinct dimension of task demand.

\subsubsection{D-Illustrate Rated Best for Supporting Multimedia Learning (LCD Questions)}  \label{sec:findings_lcd}
Linear mixed models suggested that participants rated the three LCD questions—\textit{Reducing Irrelevant Information}, \textit{Focusing on Essential Information}, and \textit{Fostering Connection between Text and Image} - differently depending on the design idea applied. Post-hoc analysis revealed that for \textit{Focusing on Essential Information}, \textit{D-Illustrate} (M = 5.50, SD = 1.15) was rated significantly higher than \textit{D-Slowdown} (M = 4.83, SD = 1.28) (t(252) = 3.34, p < .01).
For \textit{Fostering Connection between Text and Image}, \textit{D-Illustrate}  (M = 5.46, SD = 1.39) was rated significantly higher than \textit{D-Declutter} (M = 4.69, SD = 1.59) (t(252) = 3.33, p < .01). For \textit{Reducing Irrelevant Information}, there are no significant differences between the four design ideas. These findings suggest that \textit{D-Illustrate} may be especially effective in helping learners form connections between visual and textual content and focus on essential information in multimedia learning.

\subsubsection{D-Illustrate Rated Most Helpful for Reducing Task Load Demands (TLX Questions)}\label{sec:findings_nasa}

For the TLX questions, the linear mixed models indicated that participants rated the four TLX questions differently based on the design idea applied. Table \ref{tbl: nasa} shows the means and standard deviations of the ratings for both versions.
For \textit{D-Illustrate}, post-hoc analysis revealed significant reductions in mental demand (EMM = –0.71, t(30.3) = -3.07, p < .05), physical demand (EMM = –0.70, t(33.3) = -3.04, p < .05), and temporal pressure (EMM = –0.68, t(36.3) = -3.14, p < .05), along with increased learning satisfaction (EMM = 1.15, t(32.8) = -4.64, p < .001). \textit{D-Align} also led to a significant increase in learning satisfaction (EMM = 0.68, t(39.3) = -2.61, p < .05), though it showed no significant effects on mental demand, physical demand, or temporal pressure. \textit{D-Declutter} also led to a significant increase in learning satisfaction (EMM = 0.81, t(69.1) = -2.68, p < .05), though it showed no significant effects on other demands. \textit{D-Slowdown} did not produce significant changes in any of the four dimensions. Among the four design idea, \textit{D-Illustrate} had the most consistent impact, significantly reducing multiple types of cognitive demand and enhancing satisfaction, while \textit{D-Align} and \textit{D-Declutter} improved satisfaction alone.

\begin{table}[]
\resizebox{\columnwidth}{!}{%
\begin{tabular}{lcclcclcclccl}
 &
  \multicolumn{2}{c}{Mental Demand} &
   &
  \multicolumn{2}{c}{Physical Demand} &
   &
  \multicolumn{2}{c}{Temporal Pressure} &
   &
  \multicolumn{2}{c}{Learning Satisfaction} &
   \\
 &
  Original &
  Revised &
   &
  Original &
  Revised &
   &
  Original &
  Revised &
   &
  Original &
  Revised &
   \\
Design Idea &
  Mean (SD) &
  Mean (SD) &
   &
  Mean (SD) &
  Mean (SD) &
   &
  Mean (SD) &
  Mean (SD) &
   &
  Mean (SD) &
  Mean (SD) &
   \\
D-Illustrate &
  \textbf{3.09 (2.17)} &
  \textbf{2.39 (1.93)} &
  \multicolumn{1}{c}{\textbf{*}} &
  \textbf{3.01 (2.18)} &
  \textbf{2.31 (2.10)} &
  \multicolumn{1}{c}{\textbf{*}} &
  \textbf{2.96 (2.11)} &
  \textbf{2.28 (1.75)} &
  \multicolumn{1}{c}{\textbf{*}} &
  \textbf{7.75 (2.23)} &
  \textbf{8.90 (2.12)} &
  \multicolumn{1}{c}{\textbf{***}} \\
D-Align &
  2.95 (2.11) &
  2.43 (1.91) &
   &
  2.89 (2.26) &
  2.43 (2.02) &
   &
  2.88 (1.96) &
  2.35 (1.87) &
  &
  \textbf{8.06 (2.25)} &
  \textbf{8.74 (2.14)} &
  \multicolumn{1}{c}{\textbf{*}} \\
D-Declutter &
  3.04 (1.98) &
  2.63 (2.11) &
   &
  3.27 (2.46) &
  2.67 (2.30) &
   &
  2.98 (2.10) &
  2.52 (2.03) &
   &
  \textbf{7.69 (2.25)} &
  \textbf{8.50 (2.12)} &
  \multicolumn{1}{c}{*}
   \\
D-Slowdown &
  2.81 (1.74) &
  2.63 (1.92) &
   &
  2.90 (2.13) &
  2.71 (2.21) &
   &
  2.92 (1.82) &
  2.81 (2.02) &
   &
  8.29 (1.98) &
  8.23 (2.20) &
\end{tabular}%
}
\caption{Survey Results of mean and standard deviation for each TLX question item per design idea (N=16). All survey questions were based on a 10-point Likert scale, with a lower rating meaning less demand, less pressure, or less satisfaction. There are six different video clips for D-Illustrate, five for D-Align, three for D-Declutter, and three for D-Slowdown. A * denotes there are significant differences between original and revised versions at p<.05, and a *** denotes significant differences at p<.001. 
}

\label{tbl: nasa}
\Description{This table presents the findings of the means and standard deviations for original and revised versions of each TLX question item for each design idea. For Illustrate, there are significant changes in reducing mental demand, reducing physical demand, reducing temporal pressure, and increasing learning satisfaction. For D-Align, there is a significant increasing in learning satisfaction, as well as decreases in ratings for mental demand, physical demand, and temporal pressure (no significance). For D-Declutter, there is a significant change in increasing learning satisfaction, as well as decreases in mental demand, physical demand, and temporal pressure (no significance). For D-Slowdown, there are slight decreases in mental demand, physical demand, temporal pressure, and a slight decrease in learning satisfaction. 
}
\end{table}

\subsubsection{Perceived Value of Each Design Ideas}\label{sec:findings_interview} 

In the following, we present the qualitative findings of the interview results, where the participants shared their perceived values for each design idea.

\textbf{D-Illustrate: Supported Visual Understandings of Captions}
The survey results presented in Section \ref{sec:findings_nasa} suggested that D-Illustrate showed promise in all four dimensions: reducing mental demand, physical demand and temporal pressure, and increasing learning satisfaction. Furthermore, D-Illustrate showed promise in supporting two of the three learning cognitive demands: \textit{Focusing on Essential Information}, and \textit{Fostering Connection between Text and Image}.
During the interview, participants appreciated the attempt to replace irrelevant images with animated visual examples. For example, V8 commented, ``\textit{Not much need to say for [revised version]. It is just a good idea to make it visual and with what the speaker is discussing.}'' Participants also called for more parts of the video to be applied with D-Illustrate, regardless of the design idea they were presented. V12 said, ``\textit{I think that even more can be added in different parts of the video because the beginning and end started to become dull and boring.}'' Similarly, V13 mentioned, ``\textit{I think the video could be further enhanced with more application [of D-Illustrate].}''










\textbf{D-Align: Ease Visual Attention Switch between Caption and Screen.} 
D-Align primarily increased learning satisfaction, as mentioned in Section \ref{sec:findings_nasa}. 
During the interview, participants found D-Align generally helpful in optimizing when and where to look. However, three participants mentioned challenges in simultaneously focusing on captions and visual demonstrations in certain situations, as explained in the examples below. V12 suggested that on-screen text should be reduced, aligning with D-Declutter, as it could distract attention when switching between caption reading and visual viewing, ``\textit{Captions could be split into chunks and have fewer words on the screen at one time. With the graph and the words, it became a little cluttered.}'' V15 also said the added motion on-screen may distract caption reading for some DHH learners, ``\textit{The [revised] video is alright; however, the changes may confused some people while watching the captions.}'' V15 further suggested that guiding visual attention might lead to missing visual information that is not mentioned in the audio or captions but is still important,``\textit{I would like to ensure all the information is complete and not deleted while keeping it simple.}''






\textbf{D-Declutter: Less Reading Load, though sometimes Distracted Caption Reading Pace} Survey results presented in Section \ref{sec:findings_nasa} suggested that D-Declutter was rated to increase learning satisfaction. 
The participants provided overall positive feedback for D-Declutter. On the one hand, participants appreciated the attempt to highlight text using color, as V13 said, ``\textit{It was great enhance the user experience and make video content more engaging.}'' V17 also mentioned, ``\textit{I love the color text because it is clear and point.}'' On the other hand, participants found colorized text less engaging compared to visual examples, e.g., V15 said, ``\textit{I don't think [the revised version] should have that caption. The image will instead be}'' The current design was also perceived by some participants as making the captions harder to read. V8 commented, ``\textit{The red small word and the larger word are very distracting. It is difficult to tell what the speaker was referring to}'' Two participants also raised accessibility concerns for visually diverse users, such as DeafBlind learners who use braille to access video captions. Braille may not fully convey colorful information. For example, V8 noted, ``\textit{I wonder if that works for those who may have eye issues or cannot see color}.'' This highlights the need for inclusive design that considers a broader range of accessibility needs.




\textbf{D-Slowdown: Participant Not Fully Satisfied Despite Calling for Visuals to Slow Down}  
For D-Slowdown, there are no significant changes in the four dimensions for the TLX questions, as mentioned in Section \ref{sec:findings_nasa}. 
During interviews, participants commented on their reasons for the rated value in D-Slowdown. V15 said, ``\textit{I like [original] version better than the [revised] version because the [revised] version is very slow to me so that I lose track compared to the [original] version.}'' Meanwhile, in other video clips not applied with D-Slowdown, participants also mentioned the wish to make the video slower. For example, V11 suggested on a video clip applied with D-Illustrate to ``\textit{have less pictures and video or making the video longer. I notice that some of the pictures and video were too fast and I could not see and read both at the same time.}'' V3 suggested more nuanced approach to change the play speed based on different content, ``\textit{I believe the speed should correspond to the video's content: use normal speed in typical situations and slow down when something is unclear.}'' Participants compared the use of inserted pauses to uniformly slowing down the entire video and noted that “taking a break” or “a moment to breathe” felt like a more natural strategy. They expressed that evenly slowing down the video could appear unnatural, particularly in scenes involving speech or realistic scenarios.

\section{Discussion}

\subsection{Enhancing the Accessibility of Video-Lectures through Motion-Driven Designs}

Our study focuses on empowering DHH learners by altering the visual representation of learning materials through human-centered design. Although motion-driven design has been widely explored in film and media, our work is one of the first to examine motion-driven design within the context of accessible video-based learning.

\subsubsection{More Is Less: Adding Visual Content through D-Illustrate Reduces Cognitive Demand and Task Load of DHH Learners}

D-Illustrate improved the learners’ ability to focus on essential information and connect text with images. It also reduced mental demand, temporal pressure, physical demand, and increased learning satisfaction. ``Talking head clips '' were often less illustrative, reinforcing the importance of using meaningful visuals when captions carry a high reading load. Our findings echo previous work, such as signmaku \cite{chen2024towards}, which used sign-language-based commenting to visually reinforce caption content. Improving visual illustration may also benefit learners with low literacy or those more familiar with visual media formats.

\subsubsection{Guiding Visual Attention with Dynamic Cues}

D-Align overlays motion-based visual cues to direct learners’ attention to key visuals, reducing temporal pressure and improving satisfaction, especially in STEM videos featuring complex diagrams or simulations \cite{abdullah2014vstops,ge2024toward}. The disconnect between visuals and captions contributes to the challenges of split attention \cite{10.1145/3613904.3642017}. We used flashing highlights to edit the video clips following our RQ1 participants' suggestions that this is an effective way to capture attention from reading captions, similar to waving hands in Deaf culture. Future work could explore other animations such as zooming or pulsing, as well as whether visual caption styling (e.g., emotional typographies \cite{de2024caption}) could guide attention in ways similar to audio cues for hearing learners.

\subsubsection{Interpreting “Slow Down”: Speed vs. Pause}

Although participants requested to “slow down” the visuals (D-slowdown) in the formative study, our findings in the later phase suggested that the video clips applied with D-slowdown did not adequately address the corresponding video delivery challenges. However, similar requests persisted across other video clips, suggesting that we may have misinterpreted the original feedback in our video-editing process. The ASL sign for 'SLOW' may convey more than reduced speed, such as a need for deeper explanations or more time to process content. Previous work has shown that DHH individuals are highly sensitive to nuances in pacing, where changes in pause length and pausing frequency may be perceived differently from overall speed \cite{al2021different}. For example, when a human instructor is present on screen, maintaining a consistent signing speed may be appropriate, whereas explaining complex visualizations may require more frequent or extended pauses to support comprehension. Future systems could offer finer control (e.g., manual pause-to-continue options) and investigate how DHH learners define effective pacing.

\subsubsection{D-Align Needs Future Research} Some participants noted improved alignment between visuals and captions or audio, while others, particularly participants with deafness, reported no noticeable differences. The feedback from the interviews suggested that individual factors, such as residual hearing and reliance on captions, can influence how synchronization is perceived. Residual hearing refers to the remaining hearing ability a DHH individual can retain, which varies from hearing certain frequencies with hearing aids or cochlear implants to having no usable hearing at all. For example, V12 who is Hard-of-Hearing and does not lipread suggested that the revised video helped reduce the cluttered information by matching the narrator's speaking pace. She explained that she was able to ``follow'' most of the instructor's speech through residual hearing, though she couldn't hear every word clearly, ``\textit{When the diagram is made to show the parts of AR and VR, AI in the future might match all of the [visual] parts shown in the second[revised] video only introduce them when the person mentions them to reduce the amount of clutter on the screen that was shown in Version A [original].}'' Those with more residual hearing may follow spoken audio and use captions as a reference, making them more sensitive to audio-visual synchronization. In contrast, those who rely entirely on captions and visuals may not perceive sync issues unless the captions were misaligned with visual events. Given these mixed responses and limited data, D-Align warrants personalization strategies tailored to users' hearing profiles and caption usage.

\subsubsection{From Design Ideas to Scalable Motion-driven Design Principles}

Our four motion-driven design ideas reflect needs identified through participatory methods and offer new directions for accessible-first video creation. Instead of only retrofitting existing content, future work should collaborate with learning and motion designers to translate these ideas into generalizable principles. Participant ratings alone may not reflect the importance of each idea, and clip-level implementation may need refinement. The limited literature connecting motion-driven design and accessibility highlights the need for tools and workflows that support inclusive video production, potentially using multimodal generative AI such as SPiCa \cite{ning2024spica} and models of GPT-4o style \cite{xue2024longvila}.

\subsection{Enriching Multimedia Learning Theory with Diversability} 

Our work is one of the few to study Multimedia Learning Theory with DHH learners, a theory that has previously focused only on the hearing population. In our study, we found that motion-driven design can effectively support two key cognitive demands from multimedia learning theory \cite{mayer2002multimedia} - Essential and Generative Processing - with Generative Processing being the most helpful. However, Extraneous Processing is not fully supported, and further research is needed to understand its value and approaches. Generative Processing is particularly important and challenging for DHH learners, as DHH learners often need to switch between visual content and captions on-screen. This differs from the hearing population, who can simultaneously process words through ears and visual information through eyes.

Our study fills a gap in existing research, which has primarily focused on how captions and subtitles affect viewing behavior rather than the actual processing of verbal information contained in subtitles, as noted by Kruger et al.  \cite{kruger2015subtitles}. We suggest that Multimedia Learning Theory could be applied to further study DHH learners' experiences with caption design, not only in traditional 2D video but also in 3D VR/AR environments. In addition, other learning theories, such as embodied learning, may be valuable in designing and evaluating technology for DHH learners. Embodied learning emphasizes the integration of the mind with the sensorimotor systems of the body \cite{stolz2015embodied}, suggesting that cognition is deeply rooted in perception and action, and therefore it is a promising approach to the design of inclusive education.

\subsection{Methodology Reflections}
Through a formative study, we developed motion-driven design ideas based on lived experiences of DHH learners and revealed a promising participatory method: allowing visual and asynchronous contributions. This approach aligned with the strengths of the participants in visual thinking and provided a more expressive and accessible way to engage in design. Rather than compensating for hearing loss, our method centered on the Deaf culture, highlighting the visual attention and the improved perceptual strategies that many Deaf individuals develop.

However, proposing design suggestions was not always easy. Many participants had adapted to inaccessible systems and were hesitant to challenge familiar learning patterns. Some described the act of offering suggestions as burdensome, not only because articulating ideas was difficult, but also because implementing them would require effort from others that they wound not expect to see any real changes. These hesitations were shaped by educational power dynamics, where learners are rarely positioned as co-designers. To reduce these discomforts, we used speculative prompts like ``imagine you have a superpower'' and showed examples of AI-assisted generative video editing, which helped lower perceived burden and encouraged more confident and creative contributions. Looking ahead, generative AI can further reduce barriers to participation by reducing both the cognitive effort and the social cost of proposing change. Rapid, visual prototyping can help making contributions feel more feasible and welcomed.

Upon reflection, we believe the limited impact of some designs as evaluated in the evaluation phase—such as D-Slowdown—may stem from the constraints of the 2D design space, which fails to capture the temporal and spatial richness of DHH individuals' communication modes. Sign language, which a proportion of DHH individuals use, is a 3D spatial language that is based on movement, direction, and placement. This highlights the need for more immersive or multi-dimensional design methods that better support DHH individuals' communication preferences and foster participation in design-based research.

\subsection{Limitations and Future Work}

First, we proposed four motion-driven design ideas derived from a single educational video on AR technology. We do not claim these to be exhaustive. The source video - created by a technology company for desktop viewing - may have constrained the types of design ideas explored. The results may differ for videos created by college instructors or optimized for mobile platforms such as TikTok. Future work should examine additional motion-driven design possibilities using diverse video sources and presentation styles.

Second, to control variables, we only evaluated one design idea per video clip during our study. However, real-world videos often present multiple delivery challenges that may benefit from a combination of design strategies. Future research should examine how layering multiple motion-driven design ideas might better support the needs of DHH learners.

Third, all video edits were performed manually by the research team. Future work could explore how AI technologies might automate the application of motion-driven design ideas. This includes using generative AI to produce visual content, an interest expressed by the participants themselves \cite{cheng2024llm, huffman2024we}.

Fourth, in RQ2, the sequence of the video clips was kept consistent between the participants to maintain a coherent learning flow. Within each pair, the original and revised versions were shown in randomized order to reduce comparison bias. Although this setup balanced realism and control, it limited our ability to assess order effects across design ideas. Future studies should fully randomize presentation order to minimize this potential confounding, although this might be challenging in learning-related tasks.

Fifth, the number of video clips evaluated under each design idea was unbalanced due to constraints in the base video content. Despite this, our statistical model was robust to the imbalanced data, and our analysis achieved statistical power greater than 0.80. Future work could improve both balance and sample size to increase power and generalizability.

Finally, our participant sample consisted of DHH learners from a Deaf-centric university in the US, with most identified as culturally Deaf. The findings may not be generalized to all DHH individuals with different educational, linguistic, or cultural backgrounds, including people with hard-of-hearing or deaf-blindness. Our participants also raised concerns about the feasibility of standardizing AI tools given the diversity within the DHH community. Future work should involve more diverse DHH learners (e.g., from mainstream universities), and explore whether motion-driven design benefits extend to other diversability learners and hearing learners alike.





\section{Conclusion}

This mixed-method study has two-phases: first, via a formative study, DHH participants identified four key challenges in the delivery of mainstream video content and proposed the corresponding design suggestions; second, 16 DHH learners evaluated video clips that implemented four motion-driven design ideas. The findings demonstrate that these design ideas, particularly those that improved visual-audio relevance (D-Illustrate), significantly improved learners’ experiences and better supported their cognitive processing. Through our design process, we also gained a deeper understanding of how DHH learners engage with instructional video content. In particular, while previous research often emphasizes minimizing cognitive load for DHH learners by reducing content or extending time, our findings suggest that strategically adding more visual information could actually reduce cognitive workload and increase learners' satisfaction by improving clarity and temporal coherence. These results highlight the importance of further investigating the motion-driven design ideas in a wider range of educational settings and content types. In addition, the emerging capabilities of generative AI to automate and customize such motion-driven interventions present a promising direction for future research and practical implementation.

\bibliographystyle{ACM-Reference-Format}
\bibliography{sample-base}

\appendix

\clearpage
\appendix
\renewcommand{\thefigure}{A.\arabic{figure}}
\setcounter{figure}{0}

\onecolumn
\section*{Appendix}
\label{appendix}

\section{Sample Annotations} \label{appendix:designexamples}

\begin{figure}[ht]
    \centering  
    \includegraphics[width=0.8\textwidth]{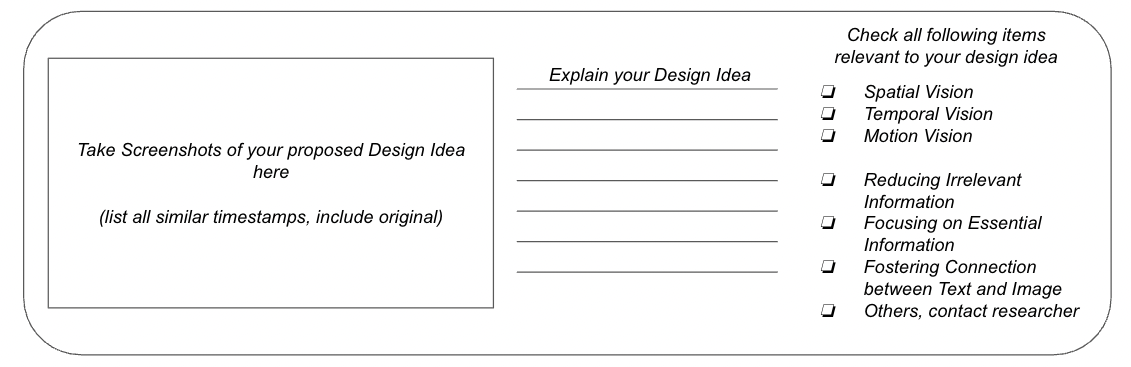}    
    \caption{Example Row from the Participant Canvas (One of Multiple Similar Entries)
    } 
    \Description{A design feedback canvas with checkboxes and text input fields. The top section includes the instruction "Check all following items relevant to your design idea" followed by a checklist: Spatial Vision, Temporal Vision, Motion Vision, Reducing Irrelevant Information, Focusing on Essential Information, Fostering Connection between Text and Image, and "Others, contact researcher." Below the checklist are two text boxes labeled "Explain your Design Idea" and "Take Screenshots of your proposed Design Idea here (list all similar timestamps, include original)."
}
    \label{fig:sample1}
\end{figure}

We present a few sample annotations collected from our participants during Phase I of the study. Each annotation includes a sketch based on the original video screenshots and design suggestions in text.

\begin{figure}[ht]
    \centering  
    \includegraphics[width=0.8\textwidth]{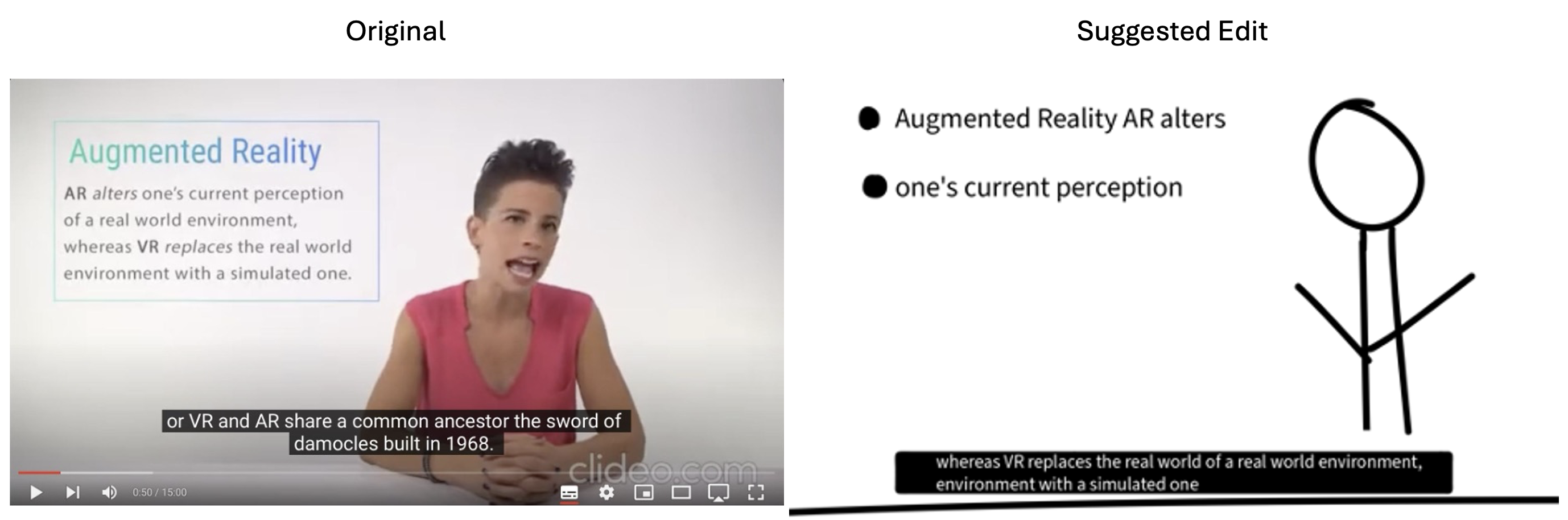}    
    \caption{
    Participant's challenge-suggestion: \textit{To enhance readability, shorten the bullet points, and display the remainder on the next clip if there are existing subtitles. This will prevent the screen from becoming too cluttered. } This informed the theme D-Declutter. Note: The participant later explained to the researcher that putting the text in the middle of the screen and potentially removing the presenter would increase focus. 
    } 
    \Description{There is one screenshot from the original video on the left and participant's suggested edit on the right. The original video shows a talking head on the right and a text box with the definition of Augmented Reality on the left with a title and four lines of text. The suggested edit replaced the long text next to talking head using bullet points.
}
    \label{fig:sample1}
\end{figure}

\begin{figure}[ht]
    \centering  
    \includegraphics[width=0.5\textwidth]{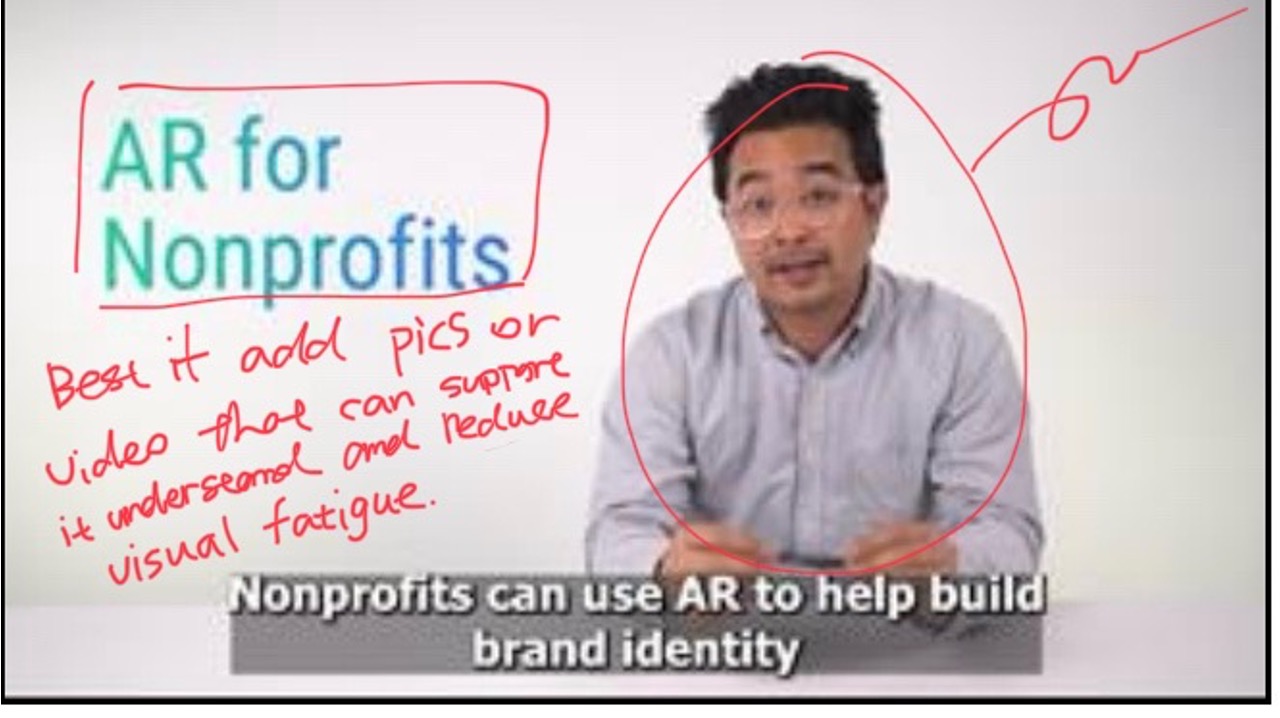}    
    \caption{Participant's challenge-suggestion: \textit{Instead of displaying the host, we can provide pictures or videos to support and help describe the topic. This will make it easier for you to understand and visualize the content being presented. }  This informed the theme D-Illustrate. Note: the talking head (male in a blue shirt) is considered irrelevant information for most participants. 
    } 
    \Description{There are several sketches on the screenshot of the original video. The original video was a talking head and text "AR for Nonprofits" next to it, with captions below. The sketch suggested to remove the talking head, and for the text, participant wrote, "Best if add pics or video that can support it understand and reduce visual fatigue."
}
    \label{fig:sample2}
\end{figure}

\begin{figure}[ht]
    \centering  
    \includegraphics[width=0.5\textwidth]{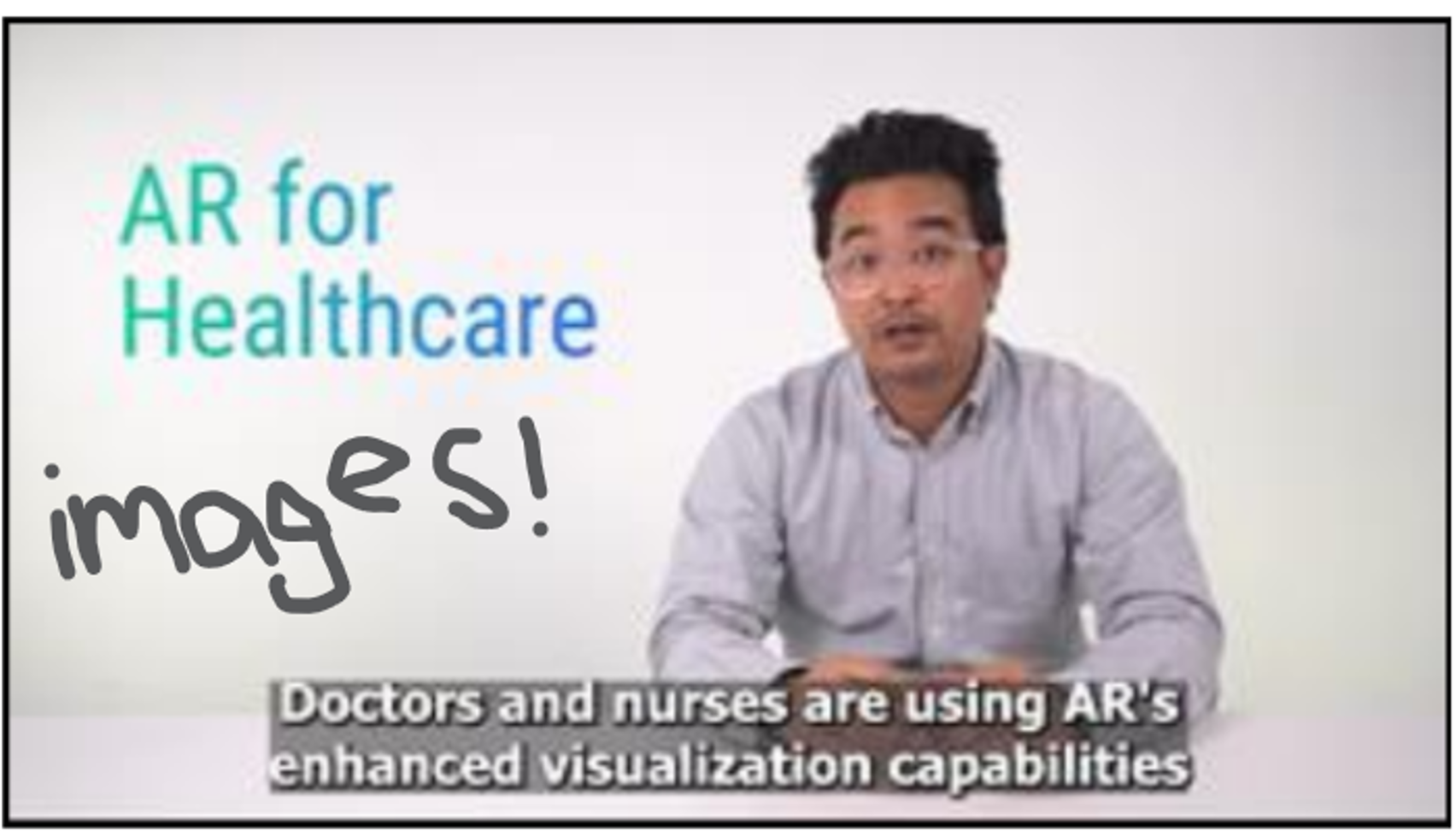}    
    \caption{Participant's challenge-suggestion: \textit{Multiple visual examples for AR for Healthcare chapter. By far the coolest/most impressive part of current AR, so doesn't make sense to not include examples of those. } This informed the theme D-Align. Note, right after the current screenshot, there are lots of visual examples that are not synchronized with the current caption.
    } 
    \Description{There is some sketch on the screenshot of the original video. The original video was a talking head and text "AR for Healthcare" next to it, with captions below. The sketch suggested to add images next to the talking head.
}
    \label{fig:sample3}
\end{figure}

\begin{figure}[ht]
    \centering  
    \includegraphics[width=0.7\textwidth]{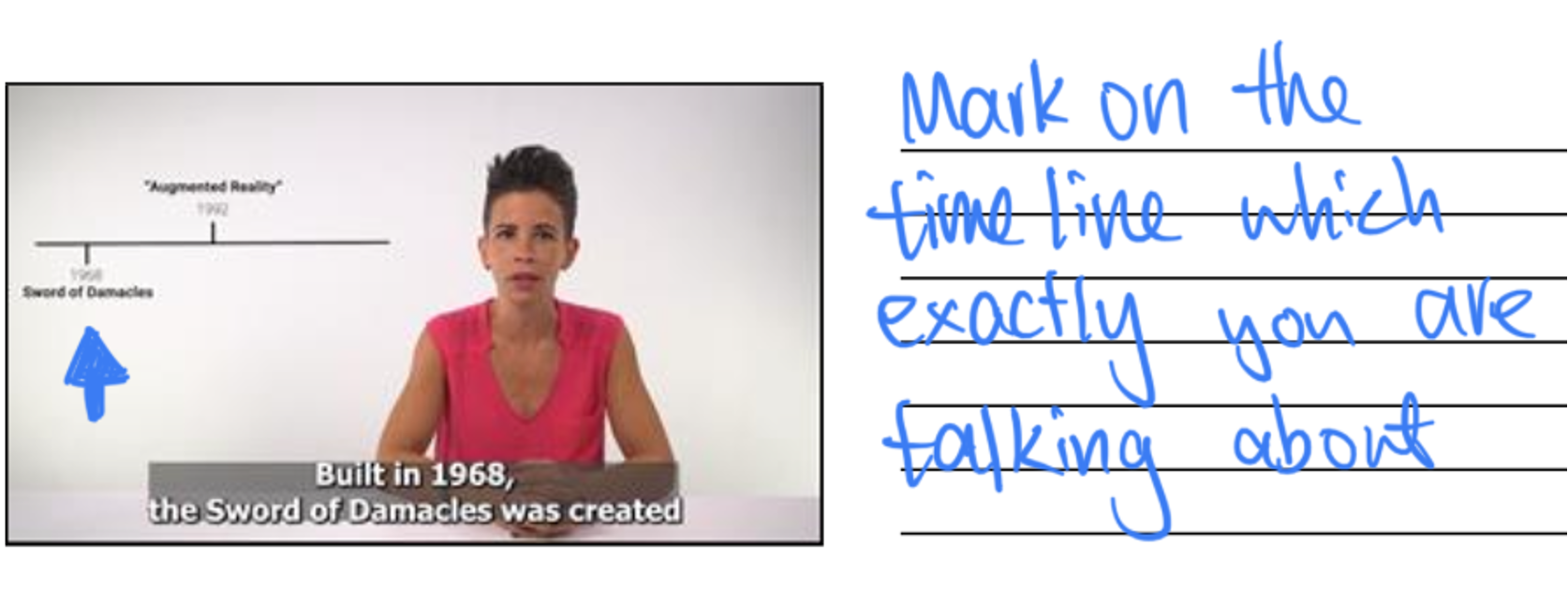}    
    \caption{Participant's challenge-suggestion: \textit{when talk about a specific point on the timeline, mark it on the timeline.} This informed the theme D-Align.
    } 
    \Description{There is some sketch on the screenshot of the original video. The original video was a talking head and a timeline of Augmented Reality next to it, with captions below. Participant's suggested edit was written on the side the screenshot: Mark on the timeline which exactly you are talking about. An arrow was added to "Sword of Damocles" highlighting the same content that was shown on the captions.
}
    \label{fig:sample4}
\end{figure}

\begin{figure}[ht]
    \centering  
    \includegraphics[width=0.5\textwidth]{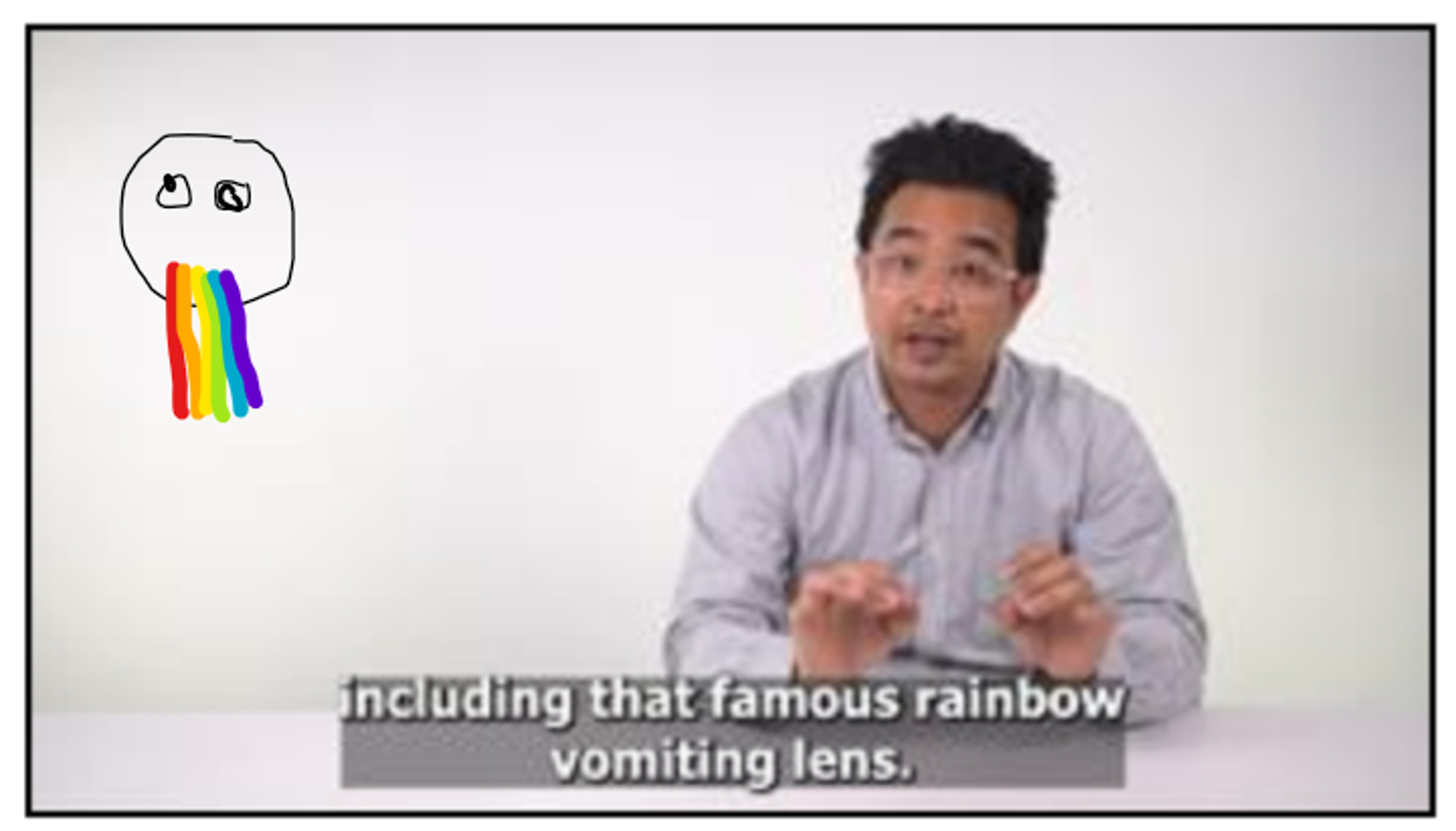}    
    \caption{Participant's challenge-suggestion: \textit{Insert example. It is possible some viewers do not know, don't want to assume/isolate audience. }  This informed the theme D-Illustrate. 
    } 
    \Description{There is some sketch on the screenshot of the original video. The original video was a talking head with captions below. The sketch was to add an image of rainbow vomiting lens that was mentioned in the caption next to the talking head.
}
    \label{fig:sample5}
\end{figure}

\begin{figure}[ht]
    \centering  
    \includegraphics[width=0.5\textwidth]{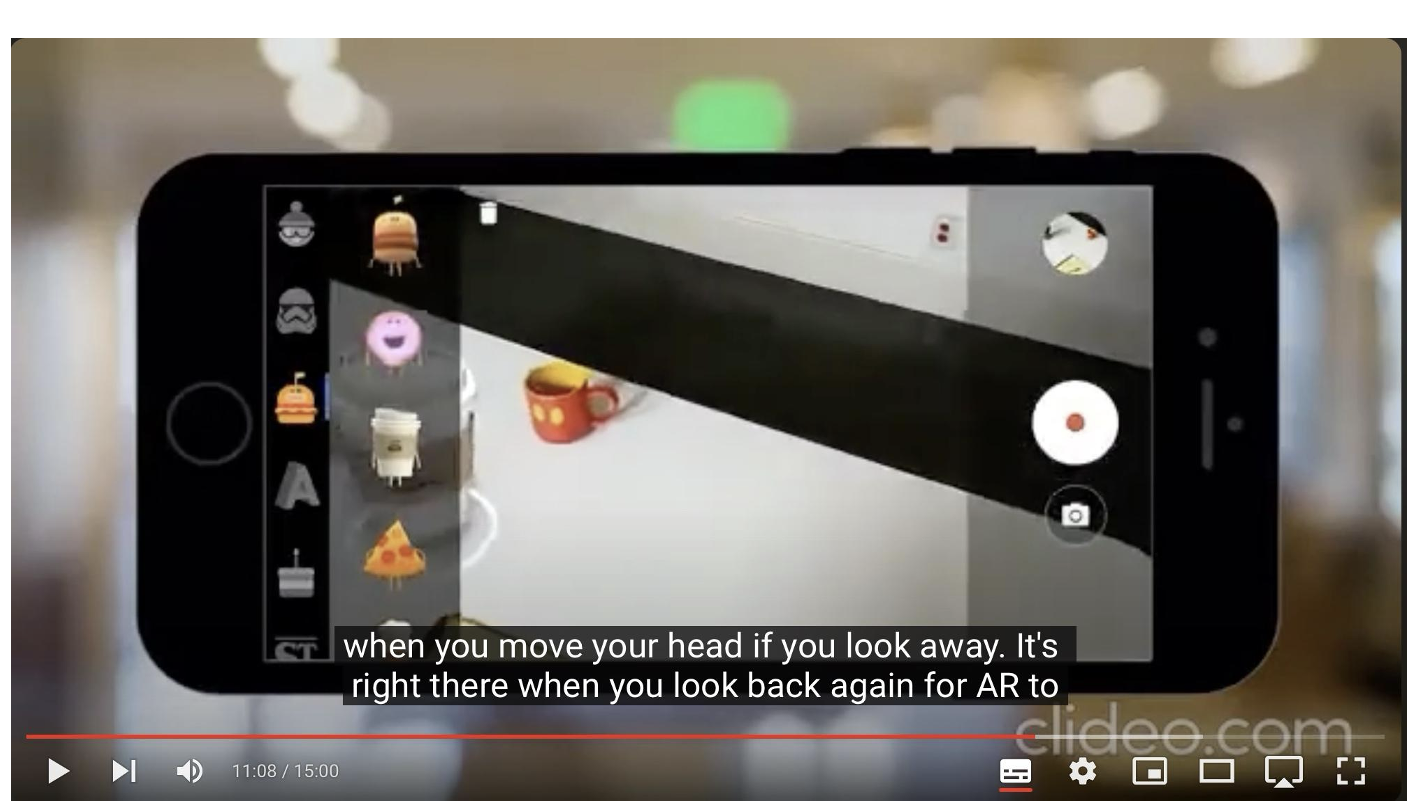}    
    \caption{Participant's challenge-suggestion: \textit{This clip moves too fast, not allowing us enough time to fully observe it. It would be better if the clip moved slowly so we could fully observe it.} This informed the theme D-Slowdown. Note: only the original video clip associated to this challenge-suggestion is shown, as there were no annotations provided with it.
    } 
    \Description{There is only the original video clip shown in this image, showing the screen of a smartphone recording a video with multiple AR objects on the screen.
}
    \label{fig:sample6}
\end{figure}

\end{document}